# Heteropolymer translocation through nanopores

Kaifu Luo[1*], Tapio Ala-Nissila[1,2], See-Chen Ying[2] and Aniket Bhattacharya[3]

[1] Laboratory of Physics, Helsinki University of Technology, P.O. Box 1100, FIN-02015 TKK, Espoo, Finland
[2] Department of Physics, Brown University, Providence, Box 1843, RI 02912-1843, U.S.A.
[3] Department of Physics, University of Central Florida, Orlando, FL 32816-2385, U.S.A.

**ABSTRACT** We investigate the translocation dynamics of heteropolymers driven through a nanopore using a constant temperature Langevin thermostat. Specifically, we consider heteropolymers consisting of two types of monomers labeled *A* and *B*, which are distinguished by the magnitude of the driving force that they experience inside the pore. From a series of studies on polymers with sequences $A_nB_{n+m}$ we identify both universal as well as sequence specific properties of the translocating chains. We find that the scaling of the average translocation time $<\tau>$ as a function of the chain length $N$ remains unaffected by the heterogeneity, while the residence time of each bead is a strong function of the sequence for short repeat units. We further discover that for a symmetric heteropolymer $A_nB_n$ of fixed length, the pattern exhibited by the residence time of the individual monomer has striking similarity with an interference pattern for an optical grating with $N/(2n)$ slits. These results are relevant for designing nanopore based sequencing techniques.

## I. Introduction

The transport of a polymer through a nanopore plays a critical role in numerous biological processes, such as DNA and RNA translocation across nuclear pores, protein transport through membrane channels, and virus injection.[1-3] Due to various potential technological applications, such as rapid DNA sequencing,[4-5] gene therapy, and controlled drug delivery,[6] the polymer translocation has become a subject of intensive experimental,[7-21] theoretical[21-43], and numerical studies.[38-55]

For a polymer threading through a nanopore, loss of available configurations due to the geometric constriction leads to an effective entropic barrier. Kasianowicz *et al.*[7] demonstrated that an electric field can drive single-stranded DNA and RNA molecules through the water-filled α-hemolysin channel and that the passage of each molecule is signaled by a blockade in the channel current. These observations can directly be used to characterize the polymer length and with further improvements the technique could be used to read off the nucleotide sequence of DNA or RNA. Other recent experiments[14] also show that an orientation asymmetry plays an important role in the translocation, due to complex interaction of DNA nucleotides with the protein nanopore. In addition to α-hemolysin channel, experiments have been done with solid state nanopore,[17-21] whose advantages are tunable pore size and stability over changes in external conditions including voltage, temperature, salinity and pH.

---

[*] Author to whom correspondence should be addressed. Electronic mail: luokaifu@yahoo.com

Theoretically, translocation dynamics was initially examined by considering equilibrium entropy of the polymer as a function of the position of the polymer through the nanopore.[23,26,28,31] This has been indicated to be inappropriate for the study of translocation dynamics.[38-44] Most recently, we have investigated both free and forced translocation using both the two-dimensional fluctuating bond (FB) model with single-segment Monte Carlo moves[42,43] and Langevin dynamic simulations.[44,45] For the free translocation,[42,44] we numerically verified that the translocation time $\tau \sim N^{1+2\nu}$, where $N$ is the chain length and $\nu$ is the Flory exponent.[56,57]. For forced translocation,[42,43] we found a crossover scaling from $\tau \sim N^{2\nu}$ for relatively short polymers to $\tau \sim N^{1+\nu}$ for longer chains.

However, most of the theoretical treatments of translocation have assumed homogenous polymer, although several experiments[12-16] show that in the real biological systems inhomogeneities in the structure and interactions between polymer and other molecules might have a significant effect on the overall dynamics. Based on the equilibrium entropy of the polymer, Muthukumar theoretically discussed some effects of sequence on translocation by considering diblock copolymers.[28] It was found that for the weak driving forces the translocation time depends on which polymer end begins the translocation process. In addition, Kafri *et al.*[30] also studied the effect of sequence heterogeneity in polymer translocation and found that the heterogeneity might lead to anomalous dynamics at some conditions.

The purpose of this paper is to investigate systematically how the pore can sense the effect of heterogeneity of the chain manifested through quantities related to its translocation dynamics. To this end, we use Langevin dynamics simulations (LD) in two dimensions (2D). We consider a coarse-grained model heteropolymers composed of two kinds of monomers labeled $A$ and $B$. Under an applied external field, the driving force for $A$ and $B$ is different. With this in mind we launch a series of investigation on chains with repeat unit $A_nB_{n+m}$ and study various properties as a function of such units. In section II, we briefly describe our model and the simulation technique. In section III, we present our results. Finally, the conclusions and discussion are in section IV.

## II. Model and method

In the simulations, the polymer chains are modeled as bead-spring chains of Lennard-Jones (LJ) particles with the finite extension nonlinear elastic (FENE) potential. Both excluded volume and van der Waals interactions between beads are modeled by a repulsive LJ potential between all bead pairs:

$$U_{LJ}(r) = \begin{cases} 4\varepsilon\left[(\sigma/r)^{12} - (\sigma/r)^{6}\right] + \varepsilon, & r \leq 2^{1/6}\sigma \\ 0, & r > 2^{1/6}\sigma, \end{cases} \qquad (1)$$

where $\sigma$ is the diameter of a bead and $\varepsilon$ is the parameter adjusting the depth of the potential.

The connectivity between the beads is modeled as a FENE spring,

$$U_{FENE}(r) = -\frac{1}{2}kR_0^2 \ln\left(1 - \frac{r^2}{R_0^2}\right), \qquad (2)$$

where $r$ is the separation between consecutive beads, $k$ is the spring constant, and $R_0$ is the maximum allowed separation between connected beads.

In the Langevin dynamics method, each bead obeys the following equation of motion:[58]

$$m\ddot{\mathbf{r}}_i = -\nabla(U_{LJ} + U_{FENE}) + \mathbf{F}_{ext} - \xi\mathbf{v}_i + \mathbf{f}_i(t), \tag{3}$$

where $m$ is the monomer's mass, $\xi$ is the friction coefficient and $\mathbf{v}_i$ is the monomer's velocity. $\mathbf{F}_{ext}$ denotes the external force due to the applied external field. $\mathbf{f}_i(t)$ is the random force and satisfies the fluctuation dissipation theorem:

$$\langle \mathbf{f}_i(t) \rangle = 0,$$
$$\langle \mathbf{f}_i(t) \cdot \mathbf{f}_j(t') \rangle = 6k_B T \xi \delta_{ij} \delta(t-t'). \tag{4}$$

The wall is described as $L$ columns of stationary particles within distance $\sigma$ from one another, which interact with the beads by the repulsive Lennard-Jones potential. The wall particle positions are not changed in the simulations. The pore is introduced in the wall by simply removing $W$ beads from each column. The external field exists only inside the pore corresponding to an external voltage applied across both ends of the pore.

In our coarse-grained model, a polymer is composed of two kinds of monomers $A$ and $B$, which may have different charges. Under the applied electric field, the driving force is proportional to the net charge on the monomer. Within the pore the external forces take the value $\mathbf{F}_{ext}=(F+\Delta)\mathbf{i}$ for monomer $A$, and $\mathbf{F}_{ext}=(F-\Delta)\mathbf{i}$ for monomer $B$, with $\mathbf{i}$ being the unit vector in the direction along the pore. For a homopolymer, $\Delta=0$. This model is not intended for a quantitative study of any real biopolymer such as DNA, where no charge difference between the monomers exists, but rather for studying the generic features induced by heterogeneity, such as scaling behavior of various physical quantities.

In our simulations, the LJ parameters $\varepsilon$ and $\sigma$ fix the system energy and length units, respectively. Time scale is given by $t_{LJ} = (m\sigma^2/\varepsilon)^{1/2}$. The parameters are $\sigma = 1$, $R_0 = 2\sigma$, $k = 7\varepsilon$, $k_B T = 1.2\varepsilon$, and reduced friction $\xi = 0.7$. In the simulations, $L = 1$ and $W = 2$ unless otherwise stated. The Langevin equation is integrated in time by a method described by Ermak and Buckholtz[59] in 2D. To create the initial configuration, the first monomer of the chain is placed in the entrance of the pore. The polymer is then let to relax to obtain an equilibrium configuration such that the first monomer position is fixed at the entrance but the other monomers are under thermal collisions described by the Langevin thermostat. In all of our simulations we did a number of runs with uncorrelated initial states and random numbers describing the random collisions. The translocation time is defined as the duration of time it takes for the chain to move through the pore in the direction of the driving force.

As to heteropolymer translocation, the reduced driving force is set as $F = 5\varepsilon/\sigma$, and $\Delta = 2.5\ \varepsilon/\sigma$ unless otherwise stated. Corresponding voltage of these parameters are in the range of real values used in the experiments.[7,11]

## III. Results and discussion

Since the primary objective is to study the effect of heterogeneous sequences it will be useful to present the simulation results into four main categories. As mentioned earlier, in our model the specific details of the beads are distinguished only when they are inside the pore where each type of bead experiences a force $F_A = (F+\Delta)$ and $F_B = (F-\Delta)$, respectively. This gross simplification is deliberately done to identify pore specific effects on the chain which is assumed to be the most dominant interaction between the two dissimilar species of monomers (notice that we do not differentiate the two species outside the pore).

## A. Effect of sequence on heteropolymer transloation with a fixed $N$

As mentioned in the introduction, one of the main objectives of studying the translocation dynamics of single molecule through nanopore is to develop methods where the pore can not only distinguish poly-nucleotides of different chain length but of a fixed length with different sequences. With that in mind we present in the following section some of the sequence specific results. Without any loss of generality here we present results for a chain of length $N = 128$, but most of our conclusions are valid for longer chains as well.

In Table 1, we show the average translocation times for multi-block copolymers with repeat units $A_mB_n$, where $m$ and $n \leq 3$. Increasing row index corresponds to increasing relative volume fraction for the species $A$. Likewise, for a given row, an increasing column index raises the volume fraction of the species $B$. Obviously, the indices $n = 0$ and $m = 0$ represent homo-polymers $A$ and $B$ respectively. We define $m + n$ as the block length, and the volume fraction of the $B$ component in the polymer $f_B = n/(m+n)$. As intuitively expected, $\tau$ increases with increasing $f_B$ because the driving force for the $B$ segments is smaller than that for the $A$ species. A three dimensional surface plot of the average translocation time is also shown in Fig. 1.

As an example, let us look at the histograms of translocation times for multiblock copolymers with repeat units $AAAB$, $AAB$, and, $AB$ more closely. Here, the values for $f_B$ are 1/4, 1/3 and 1/2, respectively. As shown in Fig. 2(a)-(c), the histograms for different polymers all obey Gaussian distribution. With increasing $f_B$, the distribution of translocation times is wider. Quantitatively, we notice that both the translocation time and the width of the distribution can be fitted with an exponential dependence on the volume fraction (Fig. 2(d)). From our data, the translocation time and the standard deviation can be approximately written as $\tau = \tau_A e^{(0.96 \pm 0.03)f_B}$ and $\sigma_\tau = \sigma_{\tau A} e^{(1.02 \pm 0.06)f_B}$, respectively, where $\tau_A$ and $\sigma_{\tau A}$ denote the corresponding translocation time and its standard deviation for homopolymer $A$. Therefore we notice that the dependence of translocation time can be expressed in a very simple form $\tau \sim \tau_A e^{f_B}$.

Interestingly, we also notice that for many short repeat units $A_mB_n$, where $m$ and $n \leq 8$, the sequence dependence of the translocation time exhibits a very rich pattern and in principle the result can be used to predict other unknown sequences as shown in Fig. 3 where the translocation times are plotted as a function of $f_B$. Here the solid line represents $\tau = \tau_A(1-f_B) + f_B\tau_B$ and the dashed line corresponds to $\tau = \tau_A e^{f_B}$. A careful look at the location of the translocation time reveals intriguing features. First, one notices that the translocation times for all sequences with either $n = 1$ and/or $m = 1$ are distributed along the curve $\tau = \tau_A e^{f_B}$. This curve ends at the pure species $A$ and $B$. For all other sequences $A_mB_n$ where $n$ and $m$ are not equal to unity, the translocation times lies close to the straight line $\tau = \tau_A(1-f_B) + f_B\tau_B$. Therefore, this type of a plot can be used to read an unknown sequence by looking at its location on the plot.

**B. Universal features of heteropolymer translocation**

In addition to exhibiting sequence dependent rich structures, the translocation of heteropolymer also exhibits certain universal aspects similar to those for the homopolymers as found in our previous studies[43,44] and by other groups[21]. Previously we observed that the translocation time and velocity of the center of mass scales as $\tau \sim N^{2\nu}$ and $v \sim N^{-\nu}$ for relatively short chains. We find that these scaling properties remain valid for heteropolymers also with arbitrary repeat unit $A_mB_n$ as shown in Fig. 4. We also checked that for chains with length longer than those shown in Fig.4, a crossover to a different exponent is clearly observed in the same manner as that for homopolymers.[43,44] This could be easily understood by noting that at a higher level of coarse-graining, the microstructure of the chain is irrelevant as far as universal scaling properties are concerned. We have checked this behavior by simulating chains of different length and repeat units $A_mB_n$.

**C. Sequence dependent features of the symmetric blocks**

Having shown the sequence dependence characteristics of the heteropolymers as a function of $f_B$, we now further analyze in detail the sequence dependent results specifically for the symmetric blocks $A_nB_n$, *i.e.* for $f_B = 0.5$. Here the block length $M = 2n$, where the minimum value of $n = 1$ for repeat unit $AB$ and its maximum value is equal to $N/2$ for a chain length $N$. Fig. 5 shows the translocation time as a function of the block length. The horizontal dotted line corresponds to $(\tau_A + \tau_B)/2$, where $\tau_A$ and $\tau_B$ are translocation times for homopolymers $A$ and $B$, respectively. We consider two different cases where monomer $A$ or monomer $B$ enters the pore first. When $A$ enters the pore first, for $M \leq 4$, $\tau$ is lower than the value $(\tau_A + \tau_B)/2$, then overshoots it with a maximum at $M \approx 16$, and finally approaches this value asymptotically for long blocks. When monomer $B$ enters the pore first, the qualitative behavior is very similar.

It is noteworthy from Fig. 5 that for $M < 16$ (for $N = 128$) the translocation time for both cases is a strong function of the block length. With increasing block length $M$ the dependence is small. This could be attributed to a decoherence effect or loss of memory for the large $A$ or $B$ segments as they go through the pore. The persistence length of the chain in the vicinity of the pore is expected to be different. Let us suppose that at a certain time the $i$-th chain is inside the pore. Evidently, one expects that the $(i - p)$-th monomer will feel this effect beyond which the chain will recover its bulk characteristics. Likewise, the $(i + q)$-th monomer ahead of this $i$-th chain will also be correlated. As a result, up to the repeat unit length $(p + q)$ the dependence will be strong. If one calculates the correlation function $\langle x(i)x(i + \delta)\rangle$ as a function of $\delta$, it will decay as a function of distance $\delta$ from the monomer $i$ located at the center of the pore.

**D. Waiting time distributions for multiblock copolymers**

Since detection of the sequence from the translocation dynamics is among the central issues in this paper, we have further investigated the increase of $\tau$ with $f_B$ in terms of the waiting (residence) time distribution time as studied previously by us for coarse-grained homopolymers[43,44]. We define the waiting or residence time of monomer $s$ as average time between the events that monomer *s* and monomer *s* + 1 exit the pore. The non-equilibrium nature of translocation has a considerable influence on this variable. We have numerically calculated the waiting times for each monomer passing through the pore.

Fig. 6 shows the waiting times for multiblock copolymers with different block size. As a reference, we also show waiting times for homopolymers *A* and *B*. As noted by our previous work[43,44] the waiting time depends strongly on the monomer position in the chain. For a homopolymer of length $N = 128$ the waiting time distribution has a maximum at a value *s* close to the middle monomer in the chain (cf. Fig. 6). We note that for longer chain, this maximum shifts towards the end of the chain.[43,44]

For the symmetric heteropolymer *AB* we notice that the residence time portrays characters of the individual monomers across the entire chain as shown in Fig. 6. An interesting result is that the residence times for the ordered heteropolymers $A_nB_n$ (for $n > 2$) exhibit "fringes" similar to optical diffraction gratings with residence time for the homopolymers *A* and *B* serving as the bounds similar to the diffraction envelope of the individual slits. Further, one immediately notices that the value of the repeat unit *n* can easily be read from the number of points between two minima, enabling to calculate length of each block. The number of peaks is exactly equal to $N/(2n)$ (128/8=16 in Fig. 6(c), 128/16=8 in Fig. 6(d), and 128/128=1 in Figs. 6(e) and (f)). Thus, if the residence time can be measured experimentally, information about the block length is immediately accessible.

The dependence of the translocation time on the block length as shown in Fig. 5 can be understood according to the waiting time distribution. As shown in Fig. 6(a), the waiting times of the monomer *B* in heteropolymer *AB* are much shorter than that for its homopolymer, while for the monomer *A* they are slightly longer. For *B* monomer the last and next monomers are *A*, which leads to less backward events, resulting in faster translocation. As a result, $\tau$ is less than $(\tau_A + \tau_B)/2$. In Fig. 6(b) and (c), for $A_2B_2$ and $A_4B_4$ heteropolymers, all the monomers in the basic block show different waiting time behavior. The first *A* and the first *B* monomers in the repeat unit show slightly longer and shorter waiting time compared with their homopolymers, respectively. However, for the other *A monomers*, their waiting times are shorter than that for homopolymer *A*, while the opposite is true for the other *B monomers*. The latter dominates the final outcome for translocation time, which leads to rapidly increase of translocation time with block size.

**IV. Summary and Conclusions**

In this work, we have investigated the dynamics of heteropolymer translocation through a nanopore driven by an external force using 2D Langevin dynamics simulations with an aim to characterize both sequence specific and universal aspects of translocation. We find that scaling exponents of the chain length dependence of the translocation time and the velocity of the center of mass are the same as that of a homopolymer. This can be easily reconciled with further coarse-graining of the chain in terms of the individual blocks of length $m + n$ for $A_mB_n$. The translocation times plotted as a function of *m* and *n* reveals novel features thus far not reported. The plots show that the translocation time $\tau = \tau(m,n)$ is unique for small values of the block $m + n$. We also notice that the average translocation time $\tau$ fall under different curves dependent on the specific patterns of the sequence. Moreover, a detailed analysis of the symmetric multiblock copolymers reveals unique and intriguing features. We find that the residence times of the individual monomers act as a fingerprint of the sequence. In particular, there is a striking similarity between the patterns observed for the distribution of the residence time for heteropolymers $A_nB_n$ of length $N$ with the interference pattern obtained from an optical diffraction grating with $N/(2n)$ slits. This mapping can possibly be extended from symmetric blocks to arbitrary repeat units $A_mB_n$, which then may lead to a better understanding of interpreting nanopore based sequence detection.

**Acknowledgments:** This work has been supported in part by the COMP Center of Excellence grant from the Academy of Finland and the TransPoly Consortium grant. We also wish to thank the Center for Scientific Computing Ltd. for allocation of computer time. A.B. gratefully acknowledges the local hospitality and travel support from Helsinki University of Technology.

| | Pure B | $A_0B_1$ | $A_0B_2$ | $A_0B_3$ |
|---|---|---|---|---|
| Pure A | | 110069±400 | 110069±400 | 110069±400 |
| $A_1B_0$ | | $A_1B_1$ | $A_1B_2$ | $A_1B_3$ |
| 41460±139 | | 66734±227 | 80610±284 | 89387±314 |
| $A_2B_0$ | | $A_2B_1$ | $A_2B_2$ | $A_2B_3$ |
| 41460±139 | | 59069±201 | 72681±258 | 81508±293 |
| $A_3B_0$ | | $A_3B_1$ | $A_3B_2$ | $A_3B_3$ |
| 41460±139 | | 55376±198 | 67625±245 | 76237±279 |

**Table 1**. The translocation times for heteropolymers of the chain length $N$=128.

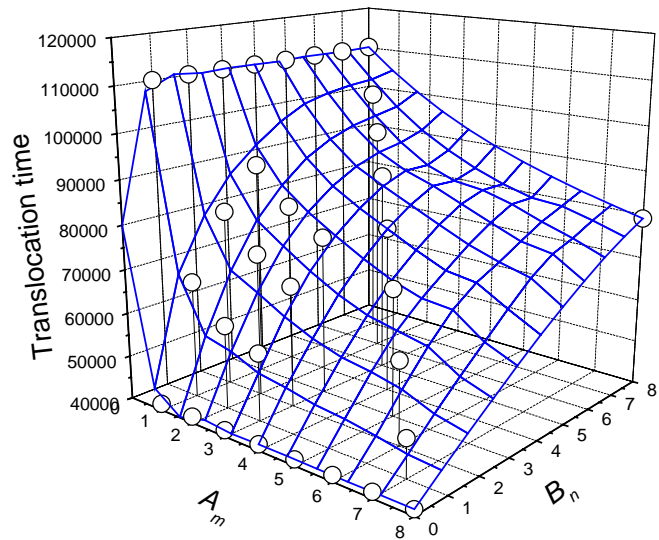

**Fig.1**. Three dimensional picture of translocation time with repeat unit.

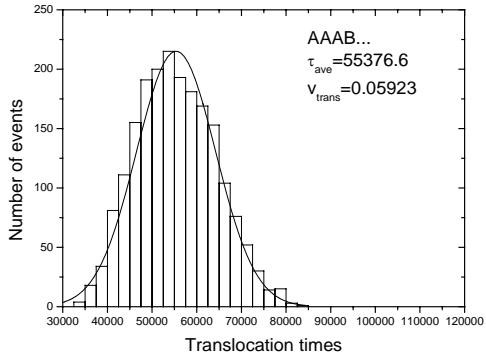

(a)

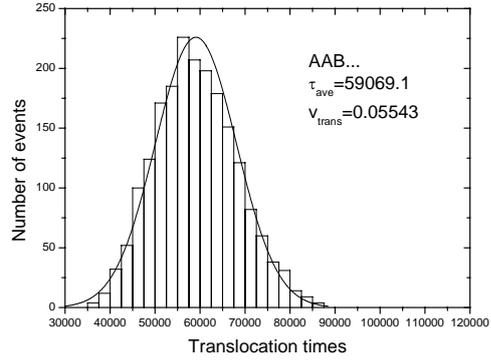

(b)

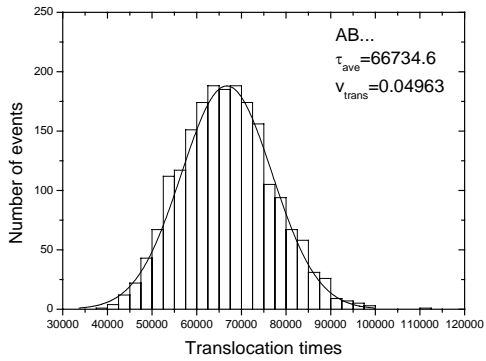

(c)

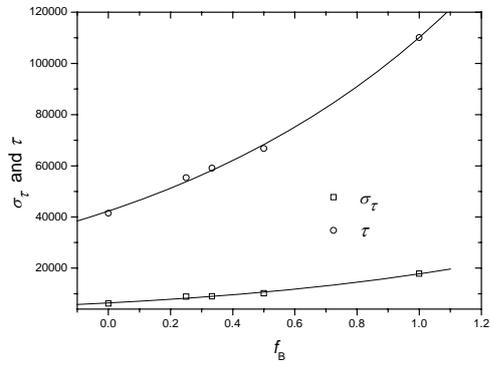

(d)

**Fig.2.** (a)-(c) Histogram of translocation times for multiblock copolymers $A_1B_1$, $A_2B_1$, and $A_3B_1$ which corresponds to $f_B$=0.5, 1/3, and 1/4 respectively. (d) Standard deviation and the translocation time as a function of the volume fraction of monomer $B$. The driving force $F$=5.0, $\Delta$=2.5 and the chain length $N$=128.

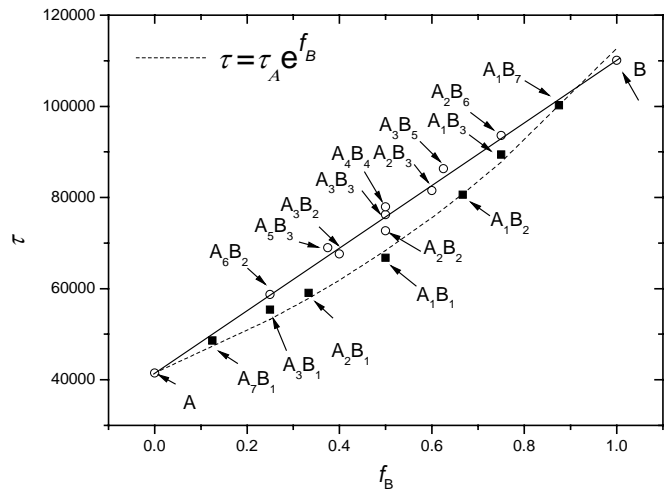

**Fig.3**. Translocation time for heteropolymers with repeat units $A_nB_m$ for different values of *m* and *n* as a function of the volume fraction of the monomer *B*.

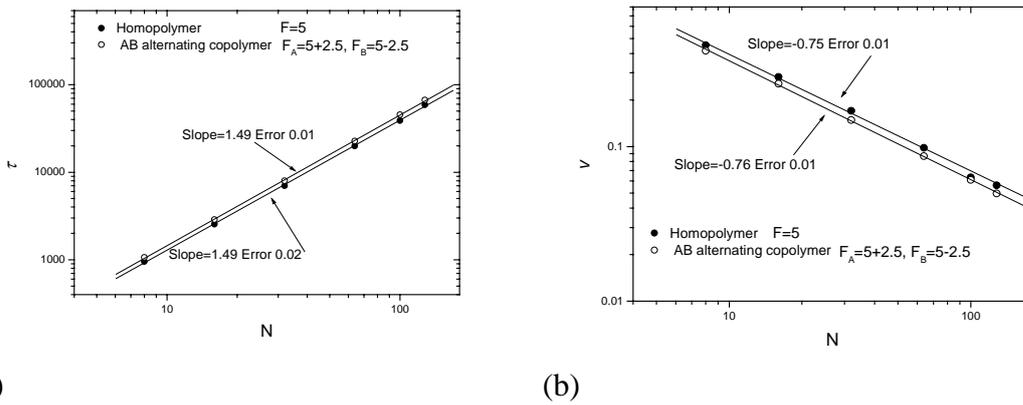

(a)                       (b)

**Fig. 4**. Scaling of forced translocation for symmetric heteropolymers $A_nB_n$ as a function of chain length: (a) scaling of translocation time, and (b) scaling of average velocity of the center of mass.

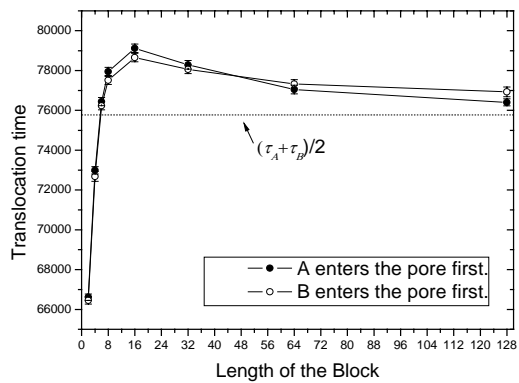

**Fig. 5**. Ttranslocation time as a function of the block length for multiblock copolymers with symmetric repeat units, $A_nB_n$. Here, the length of the block $M=2n$.

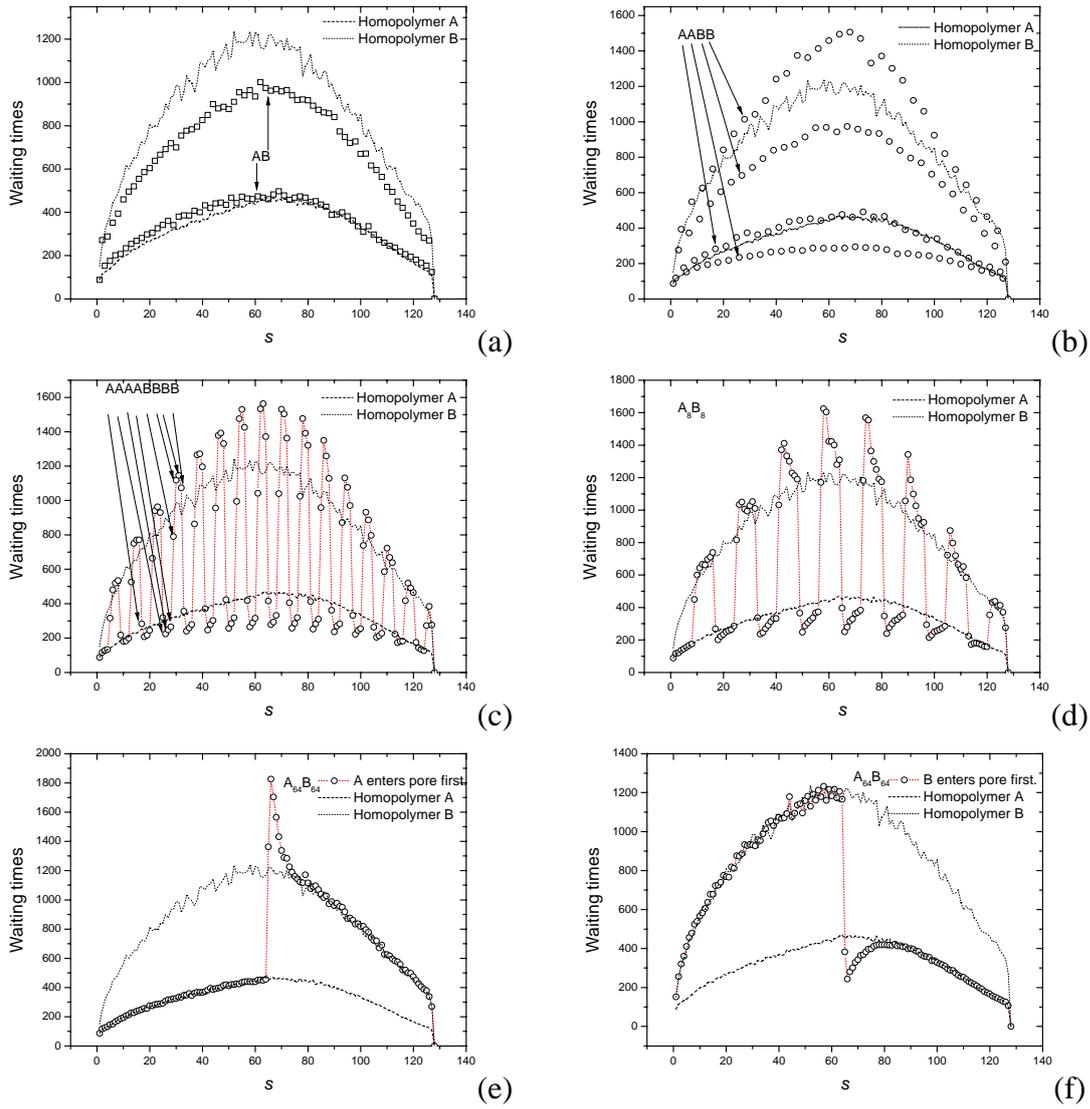

**Fig. 6**. Waiting times of all segments $s$ for multiblock copolymers with different repeat units: (a) $AB$, (b) $A_2B_2$ (c) $A_4B_4$ (d) $A_8B_8$, (e) $A_{64}B_{64}$ for the case that A enters the pore first, and (f) $A_{64}B_{64}$ for the case that B enters the pore first. The driving forces for monomers $A$ and $B$ are 7.5 and 2.5, respectively.